# Dynamics of a producer-parasite ecosystem on the brink of collapse

Andrew Chen[a,*], Alvaro Sanchez [a,b,*], Lei Dai [a] and Jeff Gore [a]

[a] Department of Physics, Massachusetts Institute of Technology, 77 Massachusetts Avenue, Cambridge MA02139, [b] Current address: The Rowland Institute at Harvard, Harvard University, 100 Edwin Land Blvd, Cambridge MA02142.

*These authors contributed equally to this work.

**Corresponding Authors:**

Jeff Gore:              gore@mit.edu

Alvaro Sanchez:         asanchez.papers@gmail.com

**ABSTRACT**

Ecosystems can undergo sudden shifts to undesirable states, but recent studies with simple single-species ecosystems have demonstrated that advance warning can be provided by the slowing down of population dynamics near a tipping point. However, it is not clear how this effect of "critical slowing down" will manifest in ecosystems with strong interactions between their components. Here we probe the dynamics of an experimental producer-parasite ecosystem as it approaches a catastrophic collapse. Surprisingly, the producer population grows in size as the environment deteriorates, highlighting that population size can be a misleading measure of ecosystem stability. By analyzing the oscillatory producer-parasite dynamics for over ~100 generations in multiple environmental conditions, we found that the collective ecosystem dynamics slows down as the tipping point is approached. Analysis of the coupled dynamics of interacting populations may therefore be necessary to provide advance warning of collapse in complex communities.

**INTRODUCTION**

Climate change and overexploitation of natural resources are altering many of the earth's ecosystems, often leading to habitat loss and species extinction. These regime shifts in ecological systems can occur without obvious warning; and once they have transpired, they may be extremely difficult to reverse, even after the agent that caused them is identified and removed (1–3). This is a consequence of the ecosystem undergoing a critical transition, in which it switches from one stable state to another. Once this happens, the feedback loops that stabilize the new state make it difficult to reverse the transition back to the previous state, leading to memory effects or hysteresis (1, 2, 4).

It has been predicted by theory that as ecosystems approach such critical transitions they may often lose resilience, making it easier for external perturbations to induce a regime shift (5). Given the negative consequences of these unwanted regime shifts, there is a desire to measure the stability of ecosystems and identify early warning indicators preceding catastrophic transitions. Theory further suggests that the loss of resilience of an ecosystem as it approaches a tipping point should be accompanied by a slowing down of the collective dynamics of the ecosystem (1, 5–9). This prediction has been confirmed in single-species laboratory microcosms, where critical slowing down and its indirect signatures (*e.g.* increases in population variability and the correlation of fluctuations) have been observed (10–12).

In parallel with the studies of simple laboratory populations, early warning indicators based on critical slowing down have been studied in complex ecosystems (2, 6, 13–16). Indeed, it is expected that sudden transitions will be common in ecological networks with multiple interacting

species(2). Theoretical analysis of concrete ecosystems with either two(6) or three(16) strongly interacting species concluded that the collapse of more complex ecosystems may also be preceded by critical slowing down - in this case manifested as the dominant eigenvalue of the community matrix approaching zero(17) (or one, for temporally discretized dynamics(18)). Encouragingly, recent experiments of exceedingly complex lake ecosystems indicate that the effects of critical slowing down may be seen by investigating the dynamics of individual species, or indirect reporters of the presence of other species(19). Nevertheless, how critical transitions take place in complex ecological networks is still poorly understood; for instance, as to how the inter-specific interactions within the ecosystem(15) affect the collective dynamics on the brink of a regime shift, or which particular species or indicators will exhibit the strongest signatures of critical slowing down. To address these questions, and to understand how early warning indicators behave in ecosystems with strong interactions between species, we set out to study the dynamics of a laboratory producer-parasite ecosystem consisting of two yeast strains with different phenotypes.

Our producer-parasite ecosystem consists of two different strains of budding yeast (*S. cerevisiae*) growing on sucrose. Budding yeast requires the enzyme invertase to break down sucrose into the simple sugars glucose and fructose. Invertase is secreted into the periplasmic space between the cell membrane and the cell wall. Due to this secretion outside of the cell, 99% of the glucose and fructose diffuse away from the cell that produced them, and can be consumed by other cells in the population(20). In our experiments, the "producer" strain expresses the SUC2 gene that encodes for invertase production; the second yeast strain, the "parasite" lacks this gene and does not contribute to the breakdown of sucrose. Thus, the parasite population exploits the producer population by consuming glucose and fructose without paying the metabolic cost associated to their production. This experimental system has been previously characterized as a model system of evolutionary game dynamics between cooperators and cheaters (20, 21), and the dynamics of a public-goods evolutionary game have been pointed out to be equivalent to a producer-parasite ecosystem (22).

**RESULTS**

**Critical collapse of an experimental producer-consumer ecosystem**

Based on prior experiments with monocultures of the producer strain in sucrose media(11), we expected our producer-parasite ecosystem to exhibit a critical collapse as the environment deteriorates. In order to test this, we grew replicate co-cultures of the two yeast strains in sucrose media, and subjected the cultures to multiple daily cycles of growth-dilution (see Methods). This naturally introduced an experimentally tunable dilution factor (DF), akin to a mortality rate, which was our control parameter for environmental deterioration (11, 21). We followed the ecosystems for nine days and measured the density of the producer and parasite populations after each growth-dilution cycle. Our observations are consistent with the presence of a tipping point in our ecosystem located around a DF of 2000: below that dilution factor the populations found a stable equilibrium, but when the DF was increased beyond this value, the populations collapsed and went extinct (Fig. 1).

In previous experimental studies in monocultures(10–12), the population size always declined with environment deterioration. In particular, this was the case for cultures of our producer strain in isolation(11). In contrast, here we found that in the face of competition with the parasite strain (whose population size does show the expected decline) the producer population increased in size as the environment deteriorated, and was maximal in the proximity of the tipping point (Fig. 1B). This increase in population size of the producer population as the environment deteriorates is a consequence of the non-linear coupled dynamics in the system (Fig. S1); at high dilution factors, the lower density of parasites leads to weaker competitive interactions and allows for less hindered growth of the producers. Species interactions may therefore lead to counter-intuitive changes at the ecosystem level as the environment deteriorates. Thus, in contrast to simple single-species ecosystems, changes in the population size of any individual species are not necessarily a reliable indicator of loss of resilience in a complex ecosystem.

**Loss of ecological resilience may be forecasted by a critical slowing down in ecosystem dynamics.**

As a warning indicator of tipping points in an ecosystem, critical slowing down is manifested in the collective dynamics of the whole ecosystem; in particular, it is characterized by the dominant eigenvalue of the linearized discrete dynamics around equilibrium (*i.e.* the interaction matrix(18)). The absolute value of the dominant eigenvalue of the interaction matrix is expected to approach $|\lambda_{dom}| = 1$ as the system approaches a critical transition(6, 23).

To observe the coupled dynamics in our population, we plotted the trajectories followed by the ecosystem on the producer-parasite phase plane (Fig. 2). Consistent with previous experiments(21), the trajectories spiraled to an equilibrium at which the two populations coexist. The observed damped oscillations are indicative of the coupled dynamics resulting from the interactions between producers and parasites. We estimated the equilibrium point in the producer-parasite phase plane (Supplementary Materials and Fig. S2), and fit the spirals to a first-order multivariate autoregressive (MAR(1)) model to obtain an estimate for the interaction matrix, from which the eigenvalues could be calculated (Supplementary Materials). As expected from the spiraling trajectories, the eigenvalues had both a real and an imaginary component. For the trajectory plotted in Fig. 2, corresponding to a dilution factor of 1333, we found $\lambda_{1,2} = |\lambda|e^{\pm i\theta}$, where $|\lambda| = 0.83 \pm 0.11$ and $\theta = 39^o \pm 6^o$. For these spiraling trajectories, the magnitude of the eigenvalue $|\lambda|$ specifies how rapidly the trajectories radially converge to the equilibrium, whereas the angle $\theta$ reflects how quickly the trajectories spiral tangentially around the equilibrium. The theory of critical slowing down predicts that $|\lambda|$ should increase approaching ~1 as the ecosystem approaches the tipping point of a critical transition. This means that as the environment deteriorates it takes longer for the two subpopulations to reach equilibrium(6).

To test these theoretical predictions, we subjected our producer-parasite ecosystems to eight different dilution factors, ranging from 50 to 1600. Thirty replicate ecosystems were tracked for each dilution factor. As expected, the absolute magnitude of the eigenvalue increased substantially as the environmental quality decreased, and approached $|\lambda| = 1$ just before population collapse (Fig. 3). Thus, our direct observation of critical slowing down in an experimental producer-parasite ecosystem lends support to the linearization approach in assessing community stability and species interactions(18), as well as using the dominant eigenvalue of the interaction matrix as a warning signal in multi-species ecosystems. We are not aware of any theoretical predictions regarding the behavior of $\theta$ (argument of eigenvalues) as the tipping point is approached, and indeed we did not observe any noticeable change in $\theta$ (Fig. S4).

**Indirect early-warning indicators of population collapse at the single-species level**

Previous analyses of natural ecosystems had indicated that individual species may exhibit behaviors consistent with critical slowing down as the ecosystem approaches a bifurcation (19). Mathematical analysis of the dynamics of a general two-species ecosystem reveals that the variability in the fluctuations of population density of each species is expected to diverge when $|\lambda|$ approaches 1 near a tipping point (18) (Supplementary Materials). Therefore, statistical indicators based on the size of fluctuations in the population density of individual species may provide indirect signatures of critical slowing down at the ecosystem level. To test this, we measured the coefficient of variation (CV) between replicates on the final day of the experiment for both the producer and parasite populations (Supplementary Materials). We found that the CV increased sharply with rising dilution factors for both producer and parasite populations (Fig. 4). Therefore, variability in the density of an individual population provides an early warning signal of the approaching tipping point in our producer-parasite ecosystem, even when the equilibrium population density increases (which is the case for our producer population).

Several other indicators of critical slowing down have been proposed theoretically and some observed experimentally in monocultures (1, 2, 10–12). They include the deterministic return time following a pulse perturbation and the lag-1 autocorrelation of the fluctuations of the population density. In our experiments with a relatively small sample size, both the deterministic return time and the lag-1 autocorrelation based on single-species time series failed to produce statistically significant increases with an increasing dilution factor (Figs. S4C,D and S5). In a previous experiment with monocultures of the producer population (11), we observed a clear increase in the coefficient of variation, lag-1 autocorrelation and return time before population collapse. Our experiments and mathematical analysis reported here suggest that certain warning indicators based on a single species may fail in the presence of interactions between species (Supplementary Materials). This highlights the importance of understanding species interactions in the application of early warning indicators to complex ecosystems; it also supports the notion that stability is a property of the entire community instead of its constituent species(24). Although properties of the whole community may not always be inferred from a single species, replicate laboratory ecosystems can help to elucidate the warning indicators which will be most useful and reliable in real-world settings.

**DISCUSSION**

Our study represents an initial characterization of population dynamics near a regime shift in an ecological network with strong inter-species interactions. It is worth noting that bifurcations in greater than two dimensions can be extraordinarily rich and complex[28]. Even in low-dimensional systems, not all bifurcations lead to critical slowing down, and regime shifts may occur with no warning [29],(27). On the other hand, recent theoretical studies have pointed out that, in some instances, critical slowing down can be detected even in the absence of any critical transitions (23, 28, 29). In spite of these potential caveats to the use of critical slowing down to forecast ecological regime shifts, our experiments demonstrate that it may be possible to observe the dominant eigenvalue of the interaction matrix increasing and approaching 1 before the collapse of a producer-parasite ecosystem, and provide support to the utility of critical slowing down as a warning signal in multi-species ecosystems (6, 24). We also recognize that there are other alternative indices based on the interaction matrix, which may prove useful in quantifying the transient dynamics (18, 30).

In many ecosystems it may not be possible to study the collective dynamics of the whole system and obtain the eigenvalues, since that would require tracking the dynamics for each of the interacting species. However, our optimism in understanding complex ecosystems come from the belief that many of them may consist of loosely coupled units with low dimensions(31). Recent experiments in a whole-lake ecosystem(19) provided evidence that some warning indicators can be observed based on time series of individual species within a multi-species ecosystem. In spite of this favorable empirical example, we caution that analyzing any single species in an ecosystem does not necessarily provide reliable warning indicators, especially when the coupling between species leads to oscillatory dynamics, as seen in this study. Even in the absence of oscillations, the projection of a species on the eigenvector corresponding to the dominant eigenvalue could in principle be small, thus making it difficult to infer the stability of an ecosystem by studying its components alone. Our study underscores the necessity of characterizing species interactions and collective dynamics when applying early warning signals to complex ecosystems. More quantitative studies in multi-species systems, especially in spatially extended contexts(32), will shed light on how we should monitor a complex ecosystem to maximize the utility of early warning indicators.

## Acknowledgments

The authors would like to thank all members of the Gore lab for useful comments and discussion. This work was supported by grant FQEB #RFP-12-07, NIH grants NIH DP2 and R00 GM085279-02, as well as grant PHY-1055154 from the NSF. The authors also received support from the Pew Foundation and the Alfred P. Sloan Foundation.

## References

1. Scheffer M et al. (2009) Early-warning signals for critical transitions. *Nature* 461:53–59.

2. Scheffer M et al. (2012) Anticipating critical transitions. *Science* 338:344–348.

3. Scheffer M, Carpenter S, Foley JA, Folke C, Walker B (2001) Catastrophic shifts in ecosystems. *Nature* 413:591–596.

4. Strogatz S (1994) *Nonlinear dynamics and chaos* (Westview Press).

5. Van Nes EH, Scheffer M (2007) Slow recovery from perturbations as a generic indicator of a nearby catastrophic shift. *Am Nat* 169:738–747.

6. Chisholm RA, Filotas E (2009) Critical slowing down as an indicator of transitions in two-species models. *J Theor Biol* 257:142–149.

7. Dakos V, van Nes EH, D'Odorico P, Scheffer M (2012) Robustness of variance and autocorrelation as indicators of critical slowing down. *Ecology* 93:264–271.

8. Dakos V et al. (2012) Methods for detecting early warnings of critical transitions in time series illustrated using simulated ecological data. *PLoS ONE* 7:e41010.

9. Carpenter SR, Brock WA, Cole JJ, Kitchell JF, Pace ML (2008) Leading indicators of trophic cascades. *Ecology Letters* 11:128–138.

10. Drake JM, Griffen BD (2010) Early warning signals of extinction in deteriorating environments. *Nature* 467:456–459.

11. Dai L, Vorselen D, Korolev KS, Gore J (2012) Generic indicators for loss of resilience before a tipping point leading to population collapse. *Science* 336:1175–1177.

12. Veraart AJ et al. (2012) Recovery rates reflect distance to a tipping point in a living system. *Nature* 481:357–359.

13. Downing AS, van Nes EH, Mooij WM, Scheffer M (2012) The resilience and resistance of an ecosystem to a collapse of diversity. *PLoS ONE* 7:e46135.

14. Biggs R, Carpenter SR, Brock WA (2009) Turning back from the brink: detecting an impending regime shift in time to avert it. *Proc Natl Acad Sci USA* 106:826–831.

15. Thrush SF et al. (2009) Forecasting the limits of resilience: integrating empirical research with theory. *Proc Biol Sci* 276:3209–3217.

16. Lade SJ, Gross T (2012) Early warning signals for critical transitions: a generalized modeling approach. *PLoS Comput Biol* 8:e1002360.

17. May RM (2001) *Stability and complexity in model ecosystems* (Princeton University Press).

18. Ives AR, Dennis B, Cottingham KL, Carpenter SR (2003) Estimating community stability and ecological interactions from time-series data. *Ecological Monographs* 73:301–330.

19. Carpenter SR et al. (2011) Early warnings of regime shifts: a whole-ecosystem experiment. *Science* 332:1079–1082.

20. Gore J, Youk H, van Oudenaarden A (2009) Snowdrift game dynamics and facultative cheating in yeast. *Nature* 459:253–256.

21. Sanchez A, Gore J (2013) Feedback between population and evolutionary dynamics determines the fate of social microbial populations. *PLoS Biology* 11:e1001547.

22. Nowak MA (2006) *Evolutionary Dynamics* (Harvard University Press).

23. Kéfi S, Dakos V, Scheffer M, Van Nes EH, Rietkerk M (2012) Early warning signals also precede non-catastrophic transitions. *Oikos*:EV1–EV8.

24. Ives AR (1995) Measuring Resilience in Stochastic Systems. *Ecological Monographs* 65:217–233.

25. Kuznetsov Y (2004) *Elements of applied bifurcation theory* (Springer). 3rd Ed.

26. Boerlijst MC, Oudman T, de Roos AM (2013) Catastrophic Collapse Can Occur without Early Warning: Examples of Silent Catastrophes in Structured Ecological Models. *PLoS ONE* 8:e62033.

27. Hastings A, Wysham DB (2010) Regime shifts in ecological systems can occur with no warning. *Ecol Lett* 13:464–472.

28. Boettiger C, Hastings A (2012) Early warning signals and the prosecutor's fallacy. *Proc Biol Sci* 279:4734–4739.

29. Boettiger C, Hastings A (2013) Tipping points: From patterns to predictions. *Nature* 493:157–158.

30. Neubert MG, Caswell H (1997) Alternatives to resilience for measuring the responses of ecological systems to perturbations. *Ecology* 78:653–665.

31. May RM (1977) Thresholds and breakpoints in ecosystems with a multiplicity of stable states. *Nature* 269:471–477.

32. Dai L, Korolev KS, Gore J (2013) Slower recovery in space before collapse of connected populations. *Nature* 496:355–358.

33. Celiker H, Gore J Competition between species can stabilize public-goods cooperation within a species. *Molecular Systems Biology* In press.

**Figures**

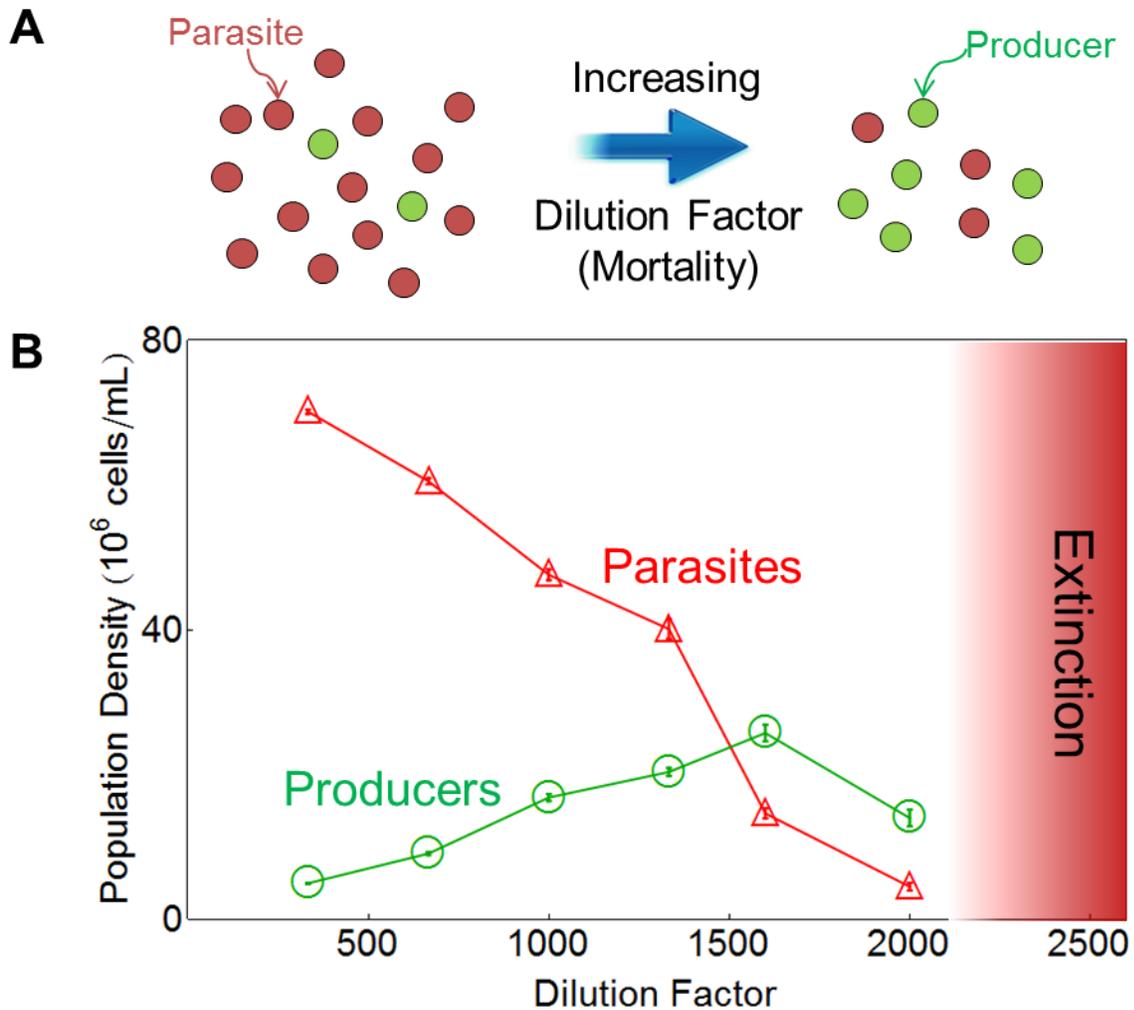

**Figure 1: Environmental deterioration leads to a surprising increase in the producer population density before collapse of the producer-parasite ecosystem.** (A) Yeast populations were grown from identical initial concentrations of producers and parasites and were subjected to a range of daily dilution factors. (B) The equilibrium population density of the producer (green) and parasite (red) populations as a function of dilution factor. Error bars represent standard error of the mean (n = 20). The producer population size increases as the environment deteriorates, highlighting that population size is not always a reliable indicator of population health.

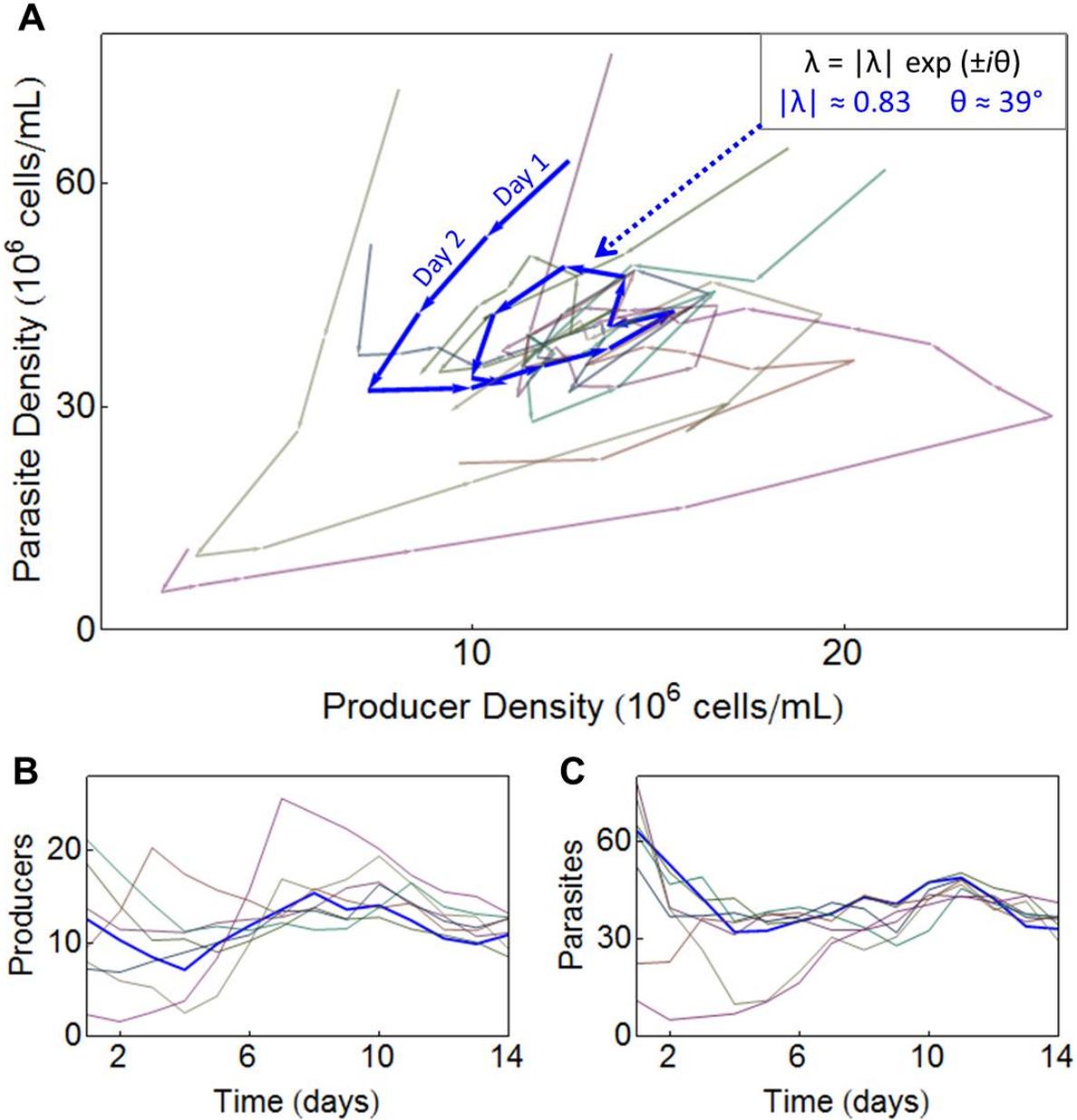

**Figure 2: The coupled spiral dynamics of the producer and parasite strains can be characterized by complex eigenvalues.** The dynamics of the population density can be visualized as trajectories in the producer-parasite phase plane (A). Replicate populations were grown from 30 different starting densities with a low daily dilution of 1333 for 14 days. Selected trajectories are colored consistently on the three plots, with remaining trajectories shown in light grey. The indicated blue trajectory has complex eigenvalues $\lambda=0.83\ e^{\pm i\,39°}$. The individual dynamics of the producers (B) and the parasites (C) are plotted as a function of time (both are expressed in units of $10^6$ cells/mL)

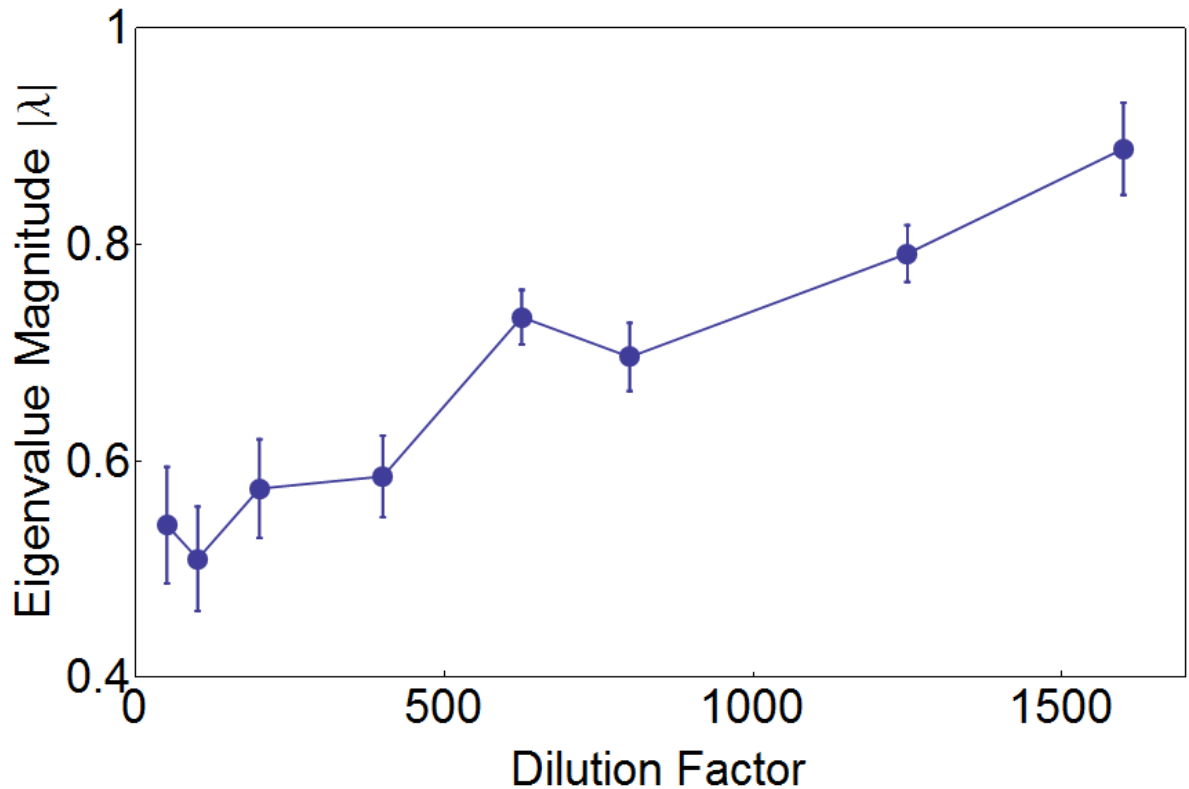

**Figure 3: Consistent with predictions from critical slowing down, the magnitude of the complex eigenvalues increases as the tipping point leading to ecosystem collapse is approached.** Thirty replicate populations with varying initial conditions were grown at each of eight different daily dilution factors. Their spiraling trajectories were analyzed to estimate their eigenvalues (generally a pair of complex conjugates, see Supplementary Materials). As predicted by theory, the magnitude of the eigenvalues increases and approaches one as the environment deteriorates. Error bars represent standard error and were obtained by bootstrapping the experimental trajectories as well as the location of the fixed point.

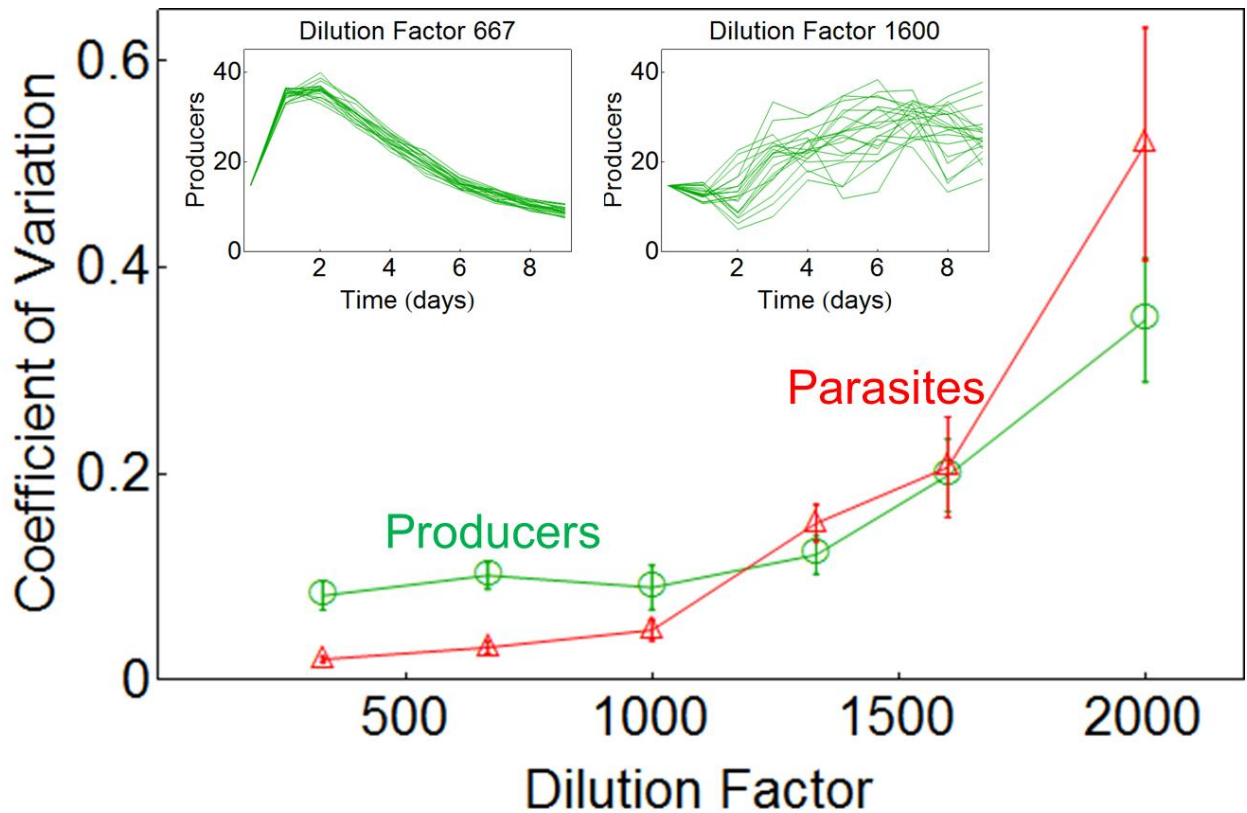

**Figure 4: Coefficient of variation increases with higher dilution factor.** The Coefficient of Variation (CV) measured for both producer (green) and parasite (red) populations as a function of dilution factor. Populations were grown from the same initial density of producers and parasites, and the coefficient of variation across the 20 replicates for each dilution factor was measured on the final day of the experiment (day 9). Inset are the time series of the producer density (measured in $10^6$ cells/mL) of each replicate at a low dilution factor (left panel) and a high dilution factor (right panel). Similar results were found for the parasite populations. Error bars are standard errors obtained by bootstrapping.

**Supplementary Materials:**

Materials and Methods

Supplementary Figures S1-S5

# Supplementary Materials

Title: Dynamics of a producer-parasite ecosystem on the brink of collapse

Andrew Chen, Alvaro Sanchez, Lei Dai and Jeff Gore

## Methods

**Experimental protocol**

*Strains*

The producer strain, JG300A(20), is derived from BY4741 strain of *Saccharomyces cerevisiae* (mating type a, EUROSCARF) that has a wild-type SUC2 gene. It constitutively expresses YFP from the *ADH1* promoter (inserted using plasmid pRS401, with a *MET17* marker), and also has a mutated *HIS3* (*his3Δ1*).

The parasite strain, JG210C (20), is a *SUC2* deletion strain (EUROSCARF Y02321, *SUC2::kanMX4*) of *Saccharomyces cerevisiae*. It constitutively expresses dTomato from the *PGK1* promoter (inserted using plasmid pRS301, with a *HIS3* marker).

*Culture growth*

The growth medium was composed of synthetic media (YNB and CSM-his; Sunrise Science, CA), 2% sucrose, 0.001% glucose, and 8 μg/mL histidine(21). Cells were cultured in 200μL of media within the 60 internal wells of a Falcon flat-bottom 96-well plate (BD Biosciences, CA) as described elsewhere(21). The plates were incubated at 30°C, while being shaken at 800 r.p.m. During incubation, the plates were covered with Parafilm as well as the plate cover. Cultures were subject to growth-dilution cycles, as described previously(21). In brief, cells were allowed to grow for 23.5 hours and then diluted by a predetermined dilution factor. The cultures were then incubated again, beginning a new day's cycle.

*Measurement of population densities*

The optical density of each culture at 620nm was measured with a Thermo Scientific Multiskan FC microplate spectrophotometer, following the end of each growth period. In addition, the saturation of the spectrophotometer at large OD values was corrected based on the following formula, which was determined previously(11):

$$Population\ Density = -6.96\times10^7 \ln\left(1-\frac{OD_m-0.038}{1.92-0.038}\right)(1.92-0.038)\frac{cells}{mL} \qquad (1)$$

To measure the fractions of each population, at the end of each growth period a portion of cells from each well (typically 5 μL, unless the low density conditions made it necessary to transfer larger volumes) was transferred to a new plate containing 200μL sterile Cellgro PBS buffer (Mediatech, VA). These samples were then analyzed at a BD LSRII-HTS flow cytometer operating in high throughput (HT) mode, where we determined the ratio between the producer and parasite strains, distinguished by their fluorescence emissions(21). Given the total density and the strain ratio, the densities of the producer and parasite populations were readily calculated.

**Data analysis**

*Filtering of Raw Data*

At higher dilution factors, some trajectories become unstable and tend towards extinction. Trajectories whose final-day population was less than $5\times10^5$ producers per mL were excluded from our analysis of fixed points and statistical indicators, as only populations that survived were relevant. For the data from the 9-day experiment (shown in Figs. 1, 4), two trajectories at a dilution factor (DF) of 2000 were also excluded because they were clearly heading towards extinction (even though the producers had not explicitly dropped below $5\times10^5$ cells/mL).

For the experiments represented in Figs. 2 and 3, populations started at different initial conditions; an additional filtering algorithm was utilized to select populations that had approached equilibrium. Trajectories whose logarithmic population densities changed less than 10% (for the experiment from Fig. 2) or 25% a day (for the experiment from Fig. 3) were defined to be in equilibrium. The mean of these populations around equilibrium on the final 3 days of the 14-day experiment or the last day of the 7-day experiment was used as an estimate of the fixed point (see Fig. S2). Due to the filter, this amounted to typically n=20 points being averaged. These estimates of the fixed points were later used to fit the Jacobian matrix (Method 1) as explained below. However, the actual trajectories analyzed for the eigenvalues were filtered by the $5\times10^5$ producers/mL criterion for survival, and the first day was dropped due to the apparent non-linear behavior far from the fixed point. Furthermore, for the calculation of lag-1 autocorrelation, only data points that had logarithmic producer and parasite densities within 25% of the estimated fixed point were used.

*Calculation of Eigenvalues*

The eigenvalues of the system are calculated by linearizing the dynamics of the trajectories around the equilibrium point. Because we observe the producer and parasite populations in discrete time intervals (once per day), we can model their motion on the producer-parasite phase plane by a difference equation:

$$\begin{pmatrix} X_{t+1} \\ Y_{t+1} \end{pmatrix} - \begin{pmatrix} X^* \\ Y^* \end{pmatrix} = J \cdot \left( \begin{pmatrix} X_t \\ Y_t \end{pmatrix} - \begin{pmatrix} X^* \\ Y^* \end{pmatrix} \right), \qquad (2)$$

where $(x_t, y_t)$ is the vector containing the producer (X) and parasite (Y) populations on day $t$. $(X^*, Y^*)$ is the equilibrium point of the trajectory, and $J$ is the 2x2 Jacobian matrix (or "interaction matrix") that describes the dynamics on the producer-parasite phase plane. Each day, this Jacobian matrix is applied to the previous day's population vector $(X_t, Y_t)$ to generate the next day's set of producer and parasite densities $(X_{t+1}, Y_{t+1})$. Notice that this is a linear transformation, which is not an exact reflection of the nonlinear dynamics for the entire phase plane. However, in the vicinity of a fixed point (as in our experiments), we may choose to retain only the first order terms of the Taylor expansion of the true transformation. The ensuing linear system is thus a reasonable approximation to our spiraling trajectories near the fixed point.

Once we determine our best estimate for the matrix $J$, we can perform an eigendecomposition to calculate its eigenvalues (here we assume that the Jacobian matrix is not defective).

$$J = P \begin{pmatrix} \lambda_1 & 0 \\ 0 & \lambda_2 \end{pmatrix} P^{-1} \quad , \qquad (3)$$

where $\lambda_1$ and $\lambda_2$ are the eigenvalues of $J$ and $P$ is the matrix whose columns are the two eigenvectors of $J$. In order to determine $J$ from our experimental data, we took the following two approaches in analyzing the trajectories, which we refer to as "Method 1" and "Method 2":

*Method 1:* The first method, is a least-squares functional minimization across the 4 parameters of $J$, given a known equilibrium point $(X^*, Y^*)$. This point is estimated for each dilution factor by the mean of the filtered data of the trajectories as shown in Fig. S2. The filter algorithm (described above) selects for data points from the last two days whose logarithmic population densities both changed by less than 25% a day.

Next, an error function $\varepsilon$ is built based on the squared difference between the true value of a trajectory's motion and its estimation given an arbitrary Jacobian matrix $J$:

$$\varepsilon = \sum_t \left[ \begin{pmatrix} X_{t+1} - X^* \\ Y_{t+1} - Y^* \end{pmatrix} - J \cdot \begin{pmatrix} X_t - X^* \\ Y_t - Y^* \end{pmatrix} \right]^2 \qquad (4)$$

This sum is expanded by inputting the trajectory data $(X_t, Y_t)$ from each day and each replicate, as well as substituting in the equilibrium point $(X^*, Y^*)$ found previously. Then, this error function of four variables is numerically minimized to locate the best estimate for $J$. This method is essentially the least-squares estimation in linear regression.

*Method 2:* A second method was also used. First, a set of possible $J$ matrices was constructed from varying ranges of six parameters: the real and imaginary parts of the eigenvalues and the eigenvectors, as well as the location of the fixed point $(X^*, Y^*)$. The matrix eigendecomposition $J$ of a given set of parameters is then used to find an error value for that set of parameters. By summing the individual errors from within each replicate and day, the total error ε is calculated from Supplementary Equation 4. The set of parameters that has the smallest total error value is deemed the best fit. Typically, around 10 values are tried for each parameter – so there are a total of a million combinations of six parameters – and this can be iterated for increased precision.

In Fig. 3, the results displayed come from Method 1 analysis. Both analysis methods yielded very similar results (Supplementary Fig. 3). There are tradeoffs in using each analysis: Method 2 does not require knowledge of the equilibrium point, but it is much more computationally intensive. Note that in Method 2, it is assumed mathematically that the eigenvalues and eigenvectors are complex, whereas Method 1 leaves open the possibility of purely real results. The fixed point ($X^*$, $Y^*$) is assumed to be real and non-negative. To attain error estimates of these parameters, the analysis was bootstrapped through a random selection of the trajectory data points and fixed point. This fixed point is randomized via a bivariate normal distribution around the estimated fixed point (see *Filtering of Raw Data*), with a standard deviation equal to the estimation's standard error (see Fig. S2).

While utilizing Method 1, we sometimes come across trajectories that give real pairs of eigenvalues, instead of complex conjugates. This generally either occurs at the lowest dilution factors (where the dynamics are too quick to be able to differentiate between real and complex eigenvalues) or at the highest dilution factors (where noise may drown out some properties of the signal). This also may be a consequence of our linearizing the dynamics of the trajectories. For the data shown in Fig. 3, only the largest dilution factor (DF 1600) had a significant number of real-eigenvalue trajectories. Thus, the mean eigenvalue magnitude was calculated by averaging the complex magnitude (in the cases of complex eigenvalues) or the dominant eigenvalue (in the cases of real eigenvalues), giving $|\lambda| = 0.88 \pm 0.04$ For comparison, when we constrained the analysis to force the eigenvalues into being complex, we found that $|\lambda| = 0.80 \pm 0.03$. In general, the difference between the dominant real eigenvalue and the complex magnitude is small when analyzed for the same trajectories. Furthermore, because constraining the eigenvalues to be complex can lead to a quantitatively worse fit, we think it is justified to analyze the eigenvalues in this mixed fashion.

*Calculation of the Return Time*

Return times were calculated by fitting an exponential to the time series of each population. Only data past the first inflexion point on each graph was used, to account for the initial overshoot (this is caused by the oscillatory behavior of the coupled producer-parasite dynamics(21)). The return time is the inverse of the exponential parameter *c* in $a + b\,e^{ct}$, (where *t* represents the time in the discrete dynamics) and is averaged across the replicate trajectories of each dilution factor. Errors were determined by taking the standard error of the return times of each individual trajectory. Further discussions on the estimation of the return time in oscillatory dynamics are in the Supplementary Text.

*Calculation of lag-1Autocorrelation*

For each population, lag-1autocorrelation (referred to as *AR(1)* later in the text) is defined as:

$$AR(1)_x = \frac{\langle (X_{t+1} - X^*)(X_t - X^*) \rangle}{\langle (X_t - X^*)^2 \rangle} = \frac{\sum_s \sum_{t=1}^{T-1} (x_{t+1}^s - \bar{x}_{t+1})(x_t^s - \bar{x}_t)}{\sum_s \sum_{t=1}^{T-1} (x_t^s - \bar{x}_t)^2} \quad (5)$$

where $x_t^s$ is the population density of replicate trajectory *s* on day *t*, $\bar{x}_t$ is the mean density of all replicates on day *t*, and *T* is the total number of days in the experiment. To estimate this from our data, we first collated the trajectories (*s*) whose producer and parasite densities were logarithmically within 25% of the mean final population size (See *Filtering of Raw Data*). From this filtered data from each day *t* of the experiment, we then calculated the mean population size across all the trajectories from each day ($\bar{x}_t$). Then, we evaluated the expected values using equation (4), applied to every two-day interval ($x_t$, $x_{t+1}$) from every replicate *s* of the experiment, from the first day (*t*=1) to the penultimate day (*t*=*T*-1). This calculation is independently performed for both the producer and parasite populations. Errors were calculated by bootstrapping both the trajectories and the two-day intervals.

**Elaboration on the experimental and data analysis procedures for main-text figures**

**Figure 1:** The data from Dilution Factors (DF) 333-2000 came from a 9-day experiment with 6 different dilution factors and about 20 replicates for each. The mean was calculated only using the population sizes from the last day (Day 9). No filtering was done except to remove obviously dying populations (see *Filtering*), which only removed two data points from DF=2000. In a separate experiment, we determined that all trajectories go extinct when DF is 2200 or higher.

**Figure 2:** This data came from 30 replicates of the same dilution factor (DF=1333) of a 14-day experiment. No trajectories were omitted from any of the graphs. The eigenvalue calculation came from Method 1 (4-parameter functional minimization), which used a fixed point determined from averaging the data present from the last 3 days of the experiment that were selected by the 10% trajectory velocity filtering algorithm described above.

**Figure 3:** The data points come from a 7-day experiment with 8 different dilution factors with 30 replicates each. Data was filtered by eliminating obviously dying trajectories. Some initial data points were omitted in the analysis due to apparent non-linear behavior of the first-day trajectories. The eigenvalues were obtained by Method 1. The fixed points used to analyze the data came from the estimates from Fig. S2B (originating from the velocity filtering algorithm applied to this 7-day experiment).

**Figure 4:** Data came from the same nine-day experiment as Fig. 1. The coefficient of variability (CV) was calculated as Standard Deviation divided over the Mean from the last day (Day 9) of the experiment. No filtering was done except to remove the two obviously dying replicates from DF=2000.

**Simulation of the experimental producer-parasite ecosystem**

We simulated a phenomenological model for our producer-parasite ecosystem, similar to the one we previously used for this system(11, 21, 33). The main assumption of this model is that the growth rate of both producers and parasites is density dependent, and given by the equation

$$\begin{pmatrix} \dot{X} \\ \dot{Y} \end{pmatrix} = \left(1 - \frac{X+Y}{K}\right) \begin{pmatrix} r_l \frac{W^n}{W^n + X^n} + r_h \frac{X^n}{W^n + X^n} & 0 \\ 0 & \gamma_l \frac{W^n}{W^n + X^n} + \gamma_h \frac{X^n}{W^n + X^n} \end{pmatrix} \begin{pmatrix} X \\ Y \end{pmatrix} \quad (6)$$

where $K$ is the combined carrying capacity of the system, and $W$ represents a threshold density of cooperators below which both the producers and the parasites grow slowly (at rates $\gamma_l$ and $r_l = (1-a)\,\gamma_l$, respectively), and above which both grow fast (at rates $\gamma_h$ and $r_h = (1+b)\,\gamma_h$ respectively). Therefore, this differential equation captures both logistic growth at high producer densities, and the Allee effect from the cooperative behavior of sucrose breakdown at low producer densities.

Notice that the differential equation is not linear, and cannot be solved analytically. After solving it numerically, we evaluated it in incremental time steps of $T=23.5$ hours (taking into account a 3 hour growth time-lag), and then the producer and parasite populations are divided by the dilution factor (mimicking the experimental dilution step).

The model is able to capture the increase in the size of the producer population as the environment deteriorated, providing further support to the idea that population sizes are not necessarily a reliable indicator of population health (Fig. S1). In our system the producer population size may increase before population collapse because the decrease in parasite population size is faster, thus reducing competition for resources. The parameters in the model are the same as we used previously(11, 21) and are summarized below:

| Parameter | Value in simulation |
|---|---|
| $\gamma_l$ | 0.31 hr$^{-1}$ |
| $\gamma_h$ | 0.47 hr$^{-1}$ |
| $W$ | 276 μL$^{-1}$ |
| $T_{lag}$ | 3 hr |
| $b$ | 0.06 |
| $a$ | 0.075 |
| $K$ | 83,341 μL$^{-1}$ |

# Supplementary Text

**Analytical derivation of the lag-1 autocorrelation (*AR*(1)) and variance of populations with species interactions**

In the framework of a first-order multivariate autoregressive model (MAR(1)) model(18, 24), we derive the asymptotic behaviors of lag-1 autocorrelation and the variance for each population in a two-species ecosystem as the magnitude of the dominant eigenvalue of the interaction matrix approaches 1. We show that for an ecosystem with two interacting species, the *AR(1)* for each species is generally not expected to be a good indicator of critical slowing down. When the species interactions lead to oscillatory dynamics (i.e. when the Jacobian has imaginary eigenvalues $\lambda_{1,2}=|\lambda|e^{\pm i\theta}$), the *AR(1)* does not approach ~*1* as the magnitude of the eigenvalue ($|\lambda|$) approaches *1*; instead *AR(1)* approaches the real part of the eigenvalue and thus may not be monotonic. Only when the trajectories do not have an oscillatory component (i.e. when the eigenvalues of the Jacobian, $\lambda_{dom}$ and $\lambda_2$, are real), the *AR(1)* approaches ~*1* as the dominant eigenvalue $\lambda_{dom}$ approaches *1*. In contrast to the *AR(1)*, we show that the variance for each population is expected to diverge as the dominant eigenvalue approaches 1, regardless of whether the eigenvalues are real or complex.

We note that the results here are only meant to show the qualitative behaviors of these two commonly used statistical indicators based on time series of a single species (partial information) when a two-species ecosystem is very close to a bifurcation associated with critical slowing down. In general, the projection of each population on the dominant eigenvector will also vary in the approach of bifurcations; thus despite their asymptotic behaviors, the indicators based on a single population may not be monotonically increasing before population collapse and this will complicate their usefulness as warning signals. However, if we also have time series of the interacting species (complete information), then we should follow our analysis on the phase plane to fit the Jacobian (Methods) and transform the variables by projecting onto the eigenvectors, which gives a more complete measure of stability of the system[5].

*MAR(1) model of a two-species ecosystem*

Let's assume a two-species ecosystem undergoing discrete dynamics as in our experiments. Let $X_t$ and $Y_t$ represent the population size for both species at day *t*. Let's also define the population size of both species at equilibrium as *X\** and *Y\**. The coupled population dynamics near equilibrium is given by the discrete equation:

$$\begin{pmatrix} X_{t+1} - X^* \\ Y_{t+1} - Y^* \end{pmatrix} = J \cdot \begin{pmatrix} X_t - X^* \\ Y_t - Y^* \end{pmatrix} + \begin{pmatrix} \xi_{x,t} \\ \xi_{y,t} \end{pmatrix}, \qquad (7)$$

where J represents the Jacobian matrix (also called "interaction matrix"). $\xi_{x,t}$ and $\xi_{y,t}$ are Gaussian while noise (we will drop the subscript *t* for convenience) and represent a random "extrinsic" noise term acting independently on each population. We assume that the extrinsic noise term for the two different populations is uncorrelated $\langle \xi_x \xi_y \rangle = 0$, and also that it is not correlated with the population size $\langle \xi_x X_t \rangle = \langle \xi_x Y_t \rangle = \langle \xi_y X_t \rangle = \langle \xi_y Y_t \rangle = 0$.

Furthermore, to simplify our equations we introduce the following notation:

$$\langle \xi_x^2 \rangle = \sigma^2, \quad x_t = X_t - X^*, \quad y_t = Y_t - Y^*,$$

where $\sigma^2$ denotes the strength of the intrinsic noise term, and $x_t$ and $y_t$ represent the deviation from equilibrium for both species.

Using this notation, equation (7) can be rewritten as:

$$\begin{pmatrix} x_{t+1} \\ y_{t+1} \end{pmatrix} = J \cdot \begin{pmatrix} x_t \\ y_t \end{pmatrix} + \begin{pmatrix} \xi_x \\ \xi_y \end{pmatrix}, \qquad (8)$$

or, in vector form:

$$\vec{x}_{t+1} = J \cdot \vec{x}_t + \vec{\xi}. \qquad (9)$$

In order to compute the covariance matrix, we first take the transpose of this equation:

$$\vec{x}_{t+1}^T = \vec{x}_t^T \cdot J^T + \vec{\xi}^T. \qquad (10)$$

Then we calculate the dot product of Equations (9) and (10).

$$\vec{x}_{t+1} \cdot \vec{x}_{t+1}^T = \left( J \cdot \vec{x}_t + \vec{\xi} \right)\left( \vec{x}_t^T \cdot J^T + \vec{\xi}^T \right) = J \cdot \vec{x}_t \cdot \vec{x}_t^T \cdot J^T + J \cdot \vec{x}_t \cdot \vec{\xi}^T + \vec{\xi} \cdot \vec{x}_t^T \cdot J^T + \vec{\xi} \cdot \vec{\xi}^T. \qquad (11)$$

In order to compute the covariance matrix, we take the time average (or, if the system is ergodic, the average over an ensemble of replicate populations at a given time) at both sides of this equation:

$$\langle \vec{x}_{t+1} \cdot \vec{x}_{t+1}^T \rangle = \langle J \cdot \vec{x}_t \cdot \vec{x}_t^T \cdot J^T \rangle + \langle J \cdot \vec{x}_t \cdot \vec{\xi}^T \rangle + \langle \vec{\xi} \cdot \vec{x}_t^T \cdot J^T \rangle + \langle \vec{\xi} \cdot \vec{\xi}^T \rangle. \qquad (12)$$

Since the system is in equilibrium, we find that:

$$\langle \vec{x}_{t+1} \cdot \vec{x}_{t+1}^T \rangle = \langle \vec{x}_t \cdot \vec{x}_t^T \rangle = \begin{pmatrix} \langle x_t^2 \rangle & \langle x_t y_t \rangle \\ \langle x_t y_t \rangle & \langle y_t^2 \rangle \end{pmatrix} = C, \qquad (13)$$

where $C$ stands for the covariance matrix. In addition, since the noise term $\vec{\xi}$ is assumed to be uncorrelated with the population size deviation $\vec{x}_t$ we also find that:

$$\left\langle J \cdot \vec{x}_t \cdot \vec{\xi}^T \right\rangle = \left\langle \vec{\xi} \cdot \vec{x}_t^T \cdot J^T \right\rangle = 0. \tag{14}$$

and by assumption $\langle \xi_x \xi_y \rangle = 0$, $\langle \xi_x^2 \rangle = \langle \xi_y^2 \rangle = \sigma^2$ (for simplicity we assumed that $\sigma^2 = \sigma_x^2 = \sigma_y^2$. In general the magnitude of extrinsic noise is different for two species, but this does not change the results on asymptotic behaviors), we get:

$$\left\langle \vec{\xi} \cdot \vec{\xi}^T \right\rangle = \sigma^2 \begin{pmatrix} 1 & 0 \\ 0 & 1 \end{pmatrix} = \sigma^2 I. \tag{15}$$

where $I$ stands for the identity matrix. Therefore, Equation (12) takes the form:

$$\left\langle \vec{x}_t \cdot \vec{x}_t^T \right\rangle = J \cdot \left\langle \vec{x}_t \cdot \vec{x}_t^T \right\rangle \cdot J^T + \left\langle \vec{\xi} \cdot \vec{\xi}^T \right\rangle \tag{16}$$

or:

$$C = J \cdot C \cdot J^T + \sigma^2 I. \tag{17}$$

In order to compute the $AR(1)$, we follow essentially the same approach: We multiply both sides of equation (8) by the transpose vector $\vec{x}_t^T$. Then, just as we did above, we take time averages at both sides of the equation. By doing this we find the following relationship between the lag-1 correlation matrix and the covariance matrix and the Jacobian:

$$B = \begin{pmatrix} \langle x_{t+1} x_t \rangle & \langle x_{t+1} y_t \rangle \\ \langle y_{t+1} x_t \rangle & \langle y_{t+1} y_t \rangle \end{pmatrix} = J \cdot C. \tag{18}$$

The $AR(1)$ for the two populations can be calculated from this matrix, since they are defined as:

$$AR(1)_x = \frac{\langle x_{t+1} x_t \rangle}{\langle x_t^2 \rangle}, \tag{19}$$

$$AR(1)_y = \frac{\langle y_{t+1} y_t \rangle}{\langle y_t^2 \rangle}. \tag{20}$$

*The variances of both populations for Jacobians with complex eigenvalues*

We have reached an equation that describes the covariance matrix as a function of the Jacobian and the extrinsic noise strength $\sigma^2$. In order to find the final closed form analytical equation for the covariances and variances from each species, as well as the *AR*(1), we need to know the Jacobian matrix. We first study the case where both eigenvalues are complex conjugate of each other. The Jacobian can be decomposed in its eigenvectors (assuming that the matrix is not defective):

$$J = P \begin{pmatrix} |\lambda|e^{i\theta} & 0 \\ 0 & |\lambda|e^{-i\theta} \end{pmatrix} P^{-1}, \qquad (21)$$

where *P* is the matrix whose columns are the eigenvectors of *J*. Since by assumption the Jacobian has complex eigenvalues, its two eigenvectors are complex conjugate of each other:

$$P = \begin{pmatrix} |u_1|e^{i\varphi_1} & |u_1|e^{-i\varphi_1} \\ |u_2|e^{i\varphi_2} & |u_2|e^{-i\varphi_2} \end{pmatrix}. \qquad (22)$$

Therefore, the Jacobian can be written in terms of the magnitude and complex argument of its eigenvalues ($|\lambda|$ and $\theta$), as well as the two parameters that characterize its eigenvectors, (*r* and *Q*, where $Q = \varphi_2 - \varphi_1$, and $r = |u_2|/|u_1|$.). By inserting equation (22) into equation (21), we get:

$$J = \begin{pmatrix} J_{11} & J_{12} \\ J_{21} & J_{22} \end{pmatrix} = |\lambda| \begin{pmatrix} Cos[\theta] + Cot[Q]Sin[\theta] & -r^{-1}Csc[Q]Sin[\theta] \\ r\,Csc[Q]Sin[\theta] & Cos[\theta] - Cot[Q]Sin[\theta] \end{pmatrix}. \qquad (23)$$

Finally, we can insert equation (23) into equation (17), and solve to obtain analytical expressions for the variances of both populations as well as their covariance. The expressions are lengthy and not particularly informative. However, we are only interested in their behavior as the absolute magnitude of the eigenvalue approaches 1, a situation corresponding to critical slowing down. We find that in the limit when $|\lambda| \to 1$:

$$\langle x_t^2 \rangle \sim \frac{\sigma^2}{1-|\lambda|} \left( \frac{Csc[Q]}{2} \right)^2 \left( \frac{1+r^2}{r^2} \right), \qquad (24)$$

$$\langle y_t^2 \rangle \sim \frac{\sigma^2}{1-|\lambda|} \left( \frac{Csc[Q]}{2} \right)^2 \left( \frac{1+r^2}{r^2} \right), \qquad (25)$$

$$\langle x_t y_t \rangle \sim \frac{\sigma^2}{1-|\lambda|} \left( \frac{Csc[Q]}{2} \frac{Cot[Q]}{2} \right) \left( \frac{1+r^2}{r} \right). \qquad (26)$$

Therefore, the variance for both populations diverges as $|\lambda| \to 1$.

*AR(1) for Jacobian matrices with complex eigenvalues*

We can compute the *AR*(1) by inserting equation (23) into equation (18). We find:

$$AR(1)_x = \frac{\langle x_{t+1} x_t \rangle}{\langle x_t^2 \rangle} = \frac{J_{11}\langle x_t^2 \rangle + J_{12}\langle x_t y_t \rangle}{\langle x_t^2 \rangle} = J_{11} + J_{12}\frac{\langle x_t y_t \rangle}{\langle x_t^2 \rangle}. \qquad (27)$$

$$AR(1)_y = \frac{\langle y_{t+1} y_t \rangle}{\langle y_t^2 \rangle} = \frac{J_{21}\langle y_t^2 \rangle + J_{22}\langle x_t y_t \rangle}{\langle y_t^2 \rangle} = J_{21} + J_{22}\frac{\langle x_t y_t \rangle}{\langle y_t^2 \rangle}. \qquad (28)$$

Now we can insert equations (23-26) into equations (26-28), and express the *AR*(1) as a function of the eigenvalues and the parameters characterizing the eigenvectors ($|\lambda|$, $\theta$, $r$, $Q$). As it was the case for the variances, the analytical equations calculated this way are long and not particularly informative. However, we can take the limit where $|\lambda| \to 1$ and find out what the behavior of the *AR*(1) in the vicinity of the critical transition. What we find is that the *AR*(1) does not approach 1, as it does for single-species ecosystems. Instead, the *AR*(1) behaves as:

$$AR(1)_x = AR(1)_y \sim |\lambda| Cos[\theta]. \qquad (29)$$

Therefore, for two-species ecosystem with strong interactions that lead to oscillatory behavior (typical of many consumer-resource ecosystems, such as predator-prey, or host-parasite) and characterized by complex eigenvalues, we conclude that the *AR*(1) for each population is not generally expected to approach 1 as the ecosystem approaches a critical transition characterized by $|\lambda| \to 1$ (i.e. critical slowing down). Thus, the failure of AR(1) in our experimental observation may be due to the fact that: 1) while the magnitude of the eigenvalue increases, the real part of the eigenvalue does not necessarily increase monotonically; 2) the relatively small sample size of our experiment tends to underestimate AR(1), based on simulations results in MAR(1) model (data not shown). Meanwhile, the variance of each population is still expected to diverge as shown in equations (24-25). Together with the experimental observations (Fig. 4 and Fig. S4B), our results suggest that variation of a single population may be a more reliable indicator than lag-1 autocorrelation.

*Variance and AR(1) for Jacobian matrices with real eigenvalues*

We can use the same procedure to show that, for Jacobian matrices with real eigenvalues, the *AR*(1) does indeed approach 1 as the dominant eigenvalue approaches 1. In addition, the variance also diverges, as it did for the case of complex eigenvalues. In order to do this, we first write down a general Jacobian matrix with real eigenvalues (we will denote the dominant eigenvalue as $\lambda_{dom}$, and the other one as $\lambda_2$). The eigendecomposition of the Jacobian on the basis of its eigenvectors is:

$$J = P \begin{pmatrix} \lambda_{dom} & 0 \\ 0 & \lambda_2 \end{pmatrix} P^{-1}. \qquad (30)$$

For convenience, we use eigenvectors of length 1. This allows us to write down the matrix *P* as:

$$P = \begin{pmatrix} Cos[\alpha] & Cos[\beta] \\ Sin[\alpha] & Sin[\beta] \end{pmatrix}, \tag{31}$$

where α and β and are the angles formed by each eigenvector with the x axis in the phase space formed by the two populations. Combining equations (30) and (31), we find the following parameterization of *J* as a function of its two eigenvalues ($\lambda_{dom}$ and $\lambda_2$), and the two angles that characterize its eigenvectors (α and β):

$$J = Csc[\alpha-\beta] \begin{pmatrix} \lambda_2 Cos[\beta]Sin[\alpha] - \lambda_{dom}Cos[\alpha]Sin[\beta] & (\lambda_{dom}-\lambda_2)Cos[\alpha]Cos[\beta] \\ -(\lambda_{dom}-\lambda_2)Sin[\alpha]Sin[\beta] & \lambda_{dom}Cos[\beta]Sin[\alpha] - \lambda_2 Cos[\alpha]Sin[\beta] \end{pmatrix} \tag{32}$$

In order to compute the *AR*(1), all we have to do is to insert equation (32) into equation (17) and find all the elements in the covariance matrix. As before, we are only interested in the behavior of these equations as $\lambda_{dom}$ approaches 1. In this limit, we find:

$$\langle x_t^2 \rangle \sim \frac{\sigma^2}{1-\lambda_{dom}} \left( \frac{Csc[\alpha-\beta]}{\sqrt{2}} \right)^2 Cos^2[\alpha], \tag{33}$$

$$\langle y_t^2 \rangle \sim \frac{\sigma^2}{1-\lambda_{dom}} \left( \frac{Csc[\alpha-\beta]}{\sqrt{2}} \right)^2 Sin^2[\alpha], \tag{34}$$

$$\langle x_t y_t \rangle \sim \frac{\sigma^2}{1-\lambda_{dom}} \left( \frac{Csc[\alpha-\beta]}{2} \right)^2 Sin[2\alpha]. \tag{35}$$

Equations (33) and (34) show that the variances of both populations diverge as $\lambda_{dom} \to 1$. We can use these equations, together with equations (27-28) to find the *AR*(1). In the same limit, when $\lambda_{dom} \to 1$, the *AR*(1) for both individual populations approaches 1. One situation that arises when dealing with real eigenvalues is that if the projection of a species on the dominant eigenvector is very small (i.e., if *Cos[α]*<<1), the variance may not diverge noticeably until the dominant eigenvalue is extremely close to 1. In conclusion, we have shown that when the eigenvalues are both real both the variance and the *AR*(1) are expected to be good indirect indicators of critical slowing down.

**Estimation of the Return Time from simulations and experiment**

In the presence of strong inter-species interactions that lead to an oscillatory component, the dynamics of both involved species will be affected by those oscillations. In the mean text we argue that the return time may be hard to estimate in these situations. Indeed, in Fig. S5 we followed what we believe would be the zero-order approach for an experimentalist trying to estimate the return time of a single species; this is, recording population size as a function of time following a perturbation or disturbance, and then fitting the observed relaxation to equilibrium to a decaying exponential to find the characteristic time of recovery. This approach has been successfully employed in previous laboratory experiments with a single species (11, 12). Furthermore, one of those experiments(11), was performed on a single-species population consisting of the same producer yeast strain we use here. Therefore, it is natural to replicate the experimental procedures that we have already established successfully for the pure parasite population, to estimate the return time in the presence of the parasite strain. As shown in Fig. S4D, we did not observe a strong increase in the return time as a function of the dilution factor. Only for the producer population we did find an increase in the return time at DF=2000, although the data is very noisy at that high DF, and we also observe a decline in the return time for the parasite population. Therefore, it appears as if the return time is not a reliable indicator of critical slowing down in our population.

In order to explain this, we resort to theory and simulations. We use a generic Jacobian matrix with complex eigenvalues to determine the population dynamics of a two-species ecosystem in following a perturbation (which we apply to a population that was previously on equilibrium by artificially increasing the size of one of the populations). In Fig. S5A we plot the response of a population to a perturbation for three different values of $|\lambda|$ (0.75, 0.95 and 0.99). We find that the dynamics of both populations are characterized by damped oscillations, as expected. As $|\lambda|$ increases, we notice that the envelope of the oscillations does indeed decay more slowly. However, the short-term dynamics (which would correspond to the naïve estimate of the return time that we applied to our experiments) is little affected by $|\lambda|$. We confirmed that this is still the case in the presence of noise (Fig. S5B), where we estimated the return time for a set of different values of $|\lambda|$ using the naïve approach.

This analysis indicates that in order to estimate the component of the return time that does increase as $|\lambda|\to 1$, we would need to observe several cycles in order to be able to estimate the decay of the envelope of the oscillations. The total number of cycles that need to be observed depend only on the argument of the complex eigenvalues $\theta$. This might lead to the requirement of very long time traces, a hurdle that cannot be overcome by increasing the sample size and observing for a shorter time. In our experiments, the envelope decay could not be seen with enough resolution for us to be able to estimate the return time.

**Supplementary Figures**

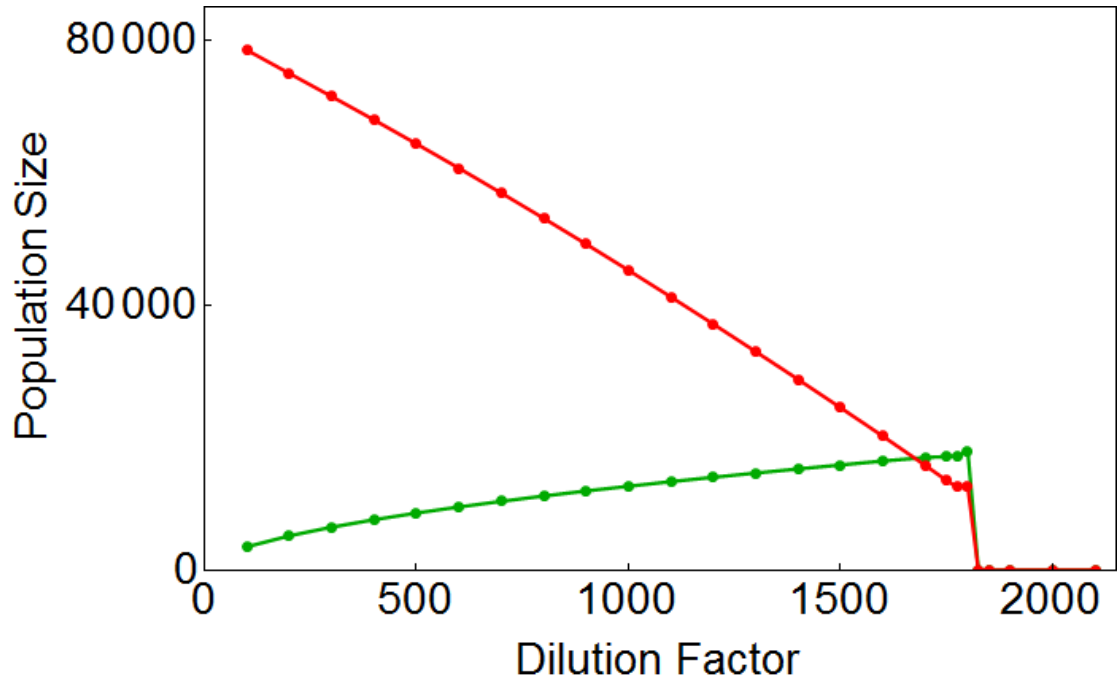

**Figure S1: Simulation of equilibrium population densities as a function of Dilution Factor has qualitative similarities to experimental results.** A noisy, simulated model of our experimental procedure was run at different dilution factors with an array of starting points. The trajectories approached an equilibrium, which was found after letting the simulation run for a prolonged period of time. The ensuing plot shows the simulated equilibrium points for the Producer (green) and Parasite (red) population densities (in cells/µL). By dilution factors above 1800, all populations go extinct, regardless of the location of the starting points.

A

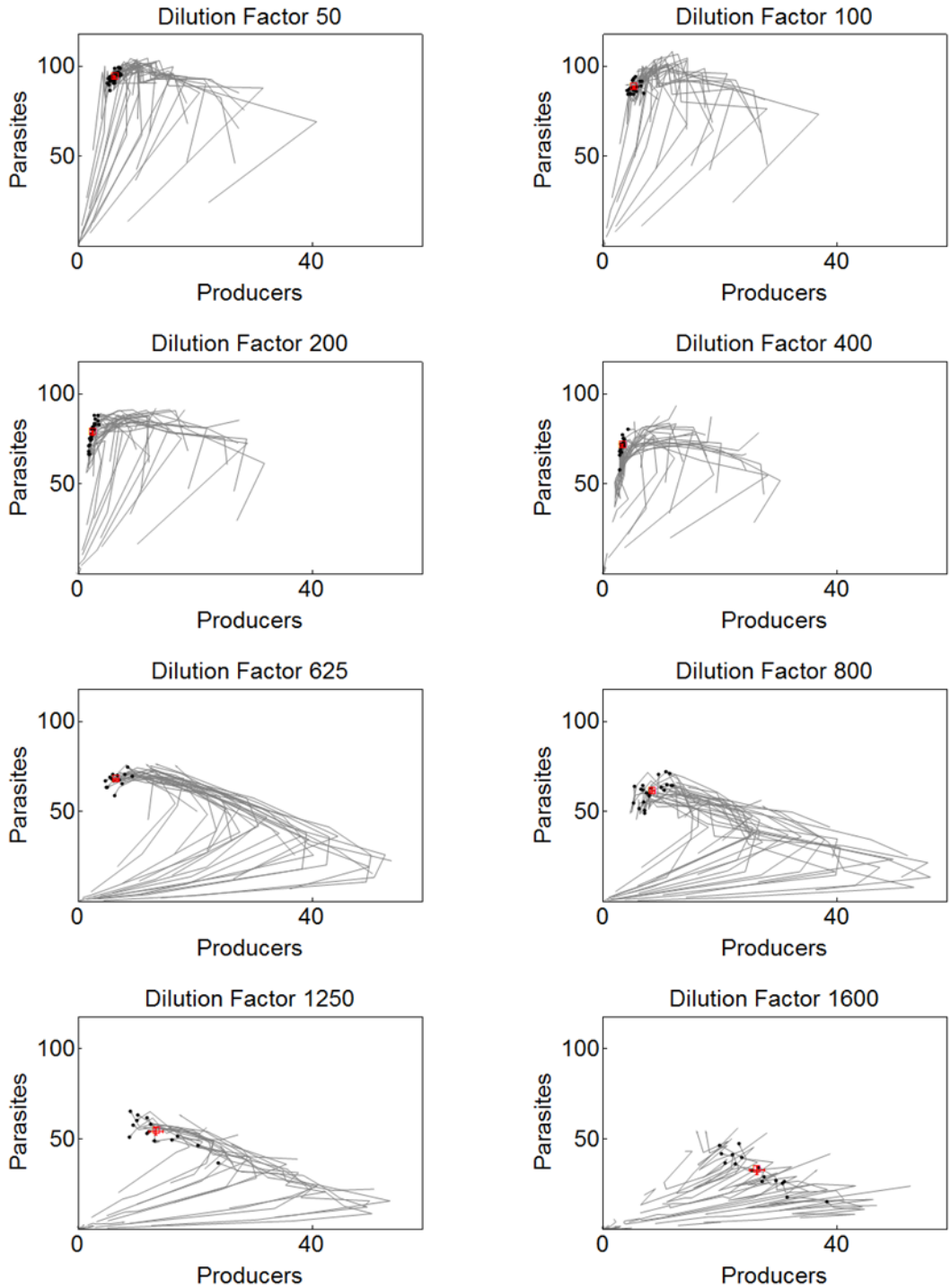

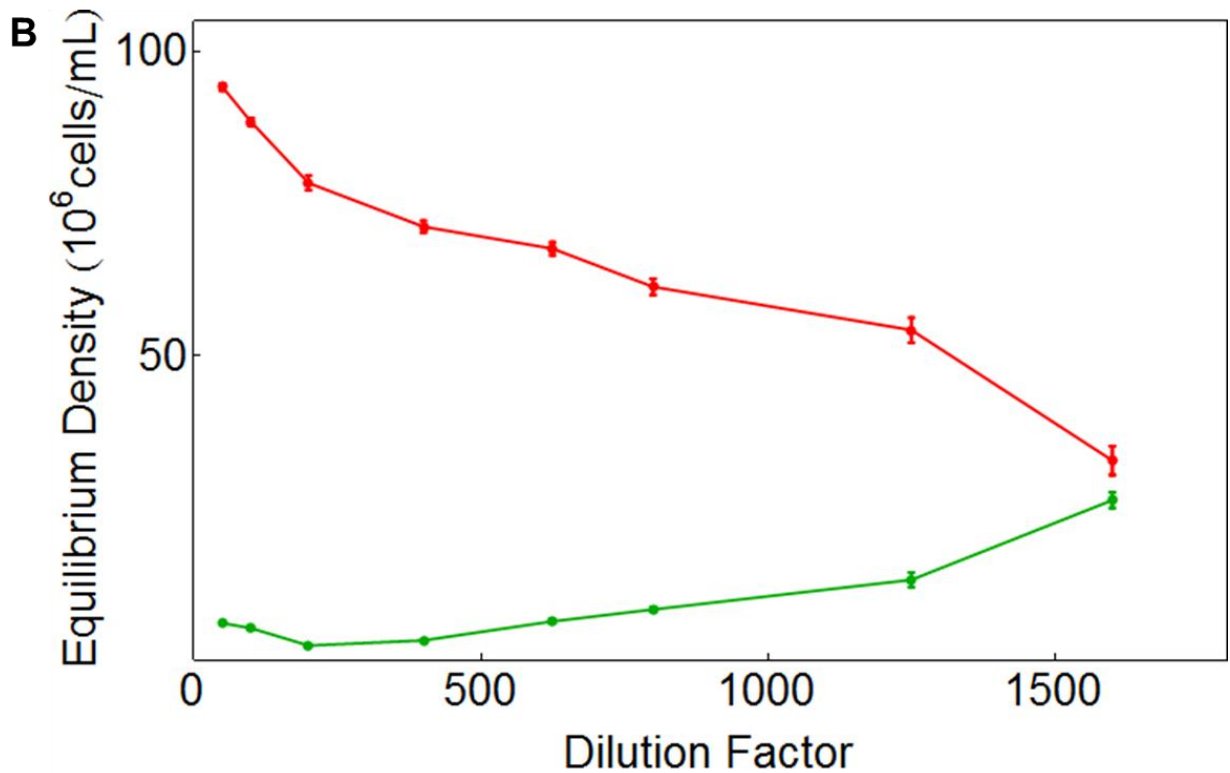

**Figure S2: Spiraling trajectory graphs, filtering, and calculation of equilibrium points. (A)** The spiraling trajectories of all dilution factors of the 7-day experiment represented in Fig. 3, plotted in gray on the same Producer – Parasite plane. Population densities are all expressed in units of $10^6$ cells/mL. Black points denote the final state of those trajectories that survived the filtering algorithm described in Methods, and which are thus considered to be in equilibrium. The red squares represent the mean of these filtered points with accompanying standard errors – which was further used as an estimate of the equilibrium point of the spirals. **(B)** These were plotted as a function of the dilution factor for both the Producers (green) and Parasites (red).

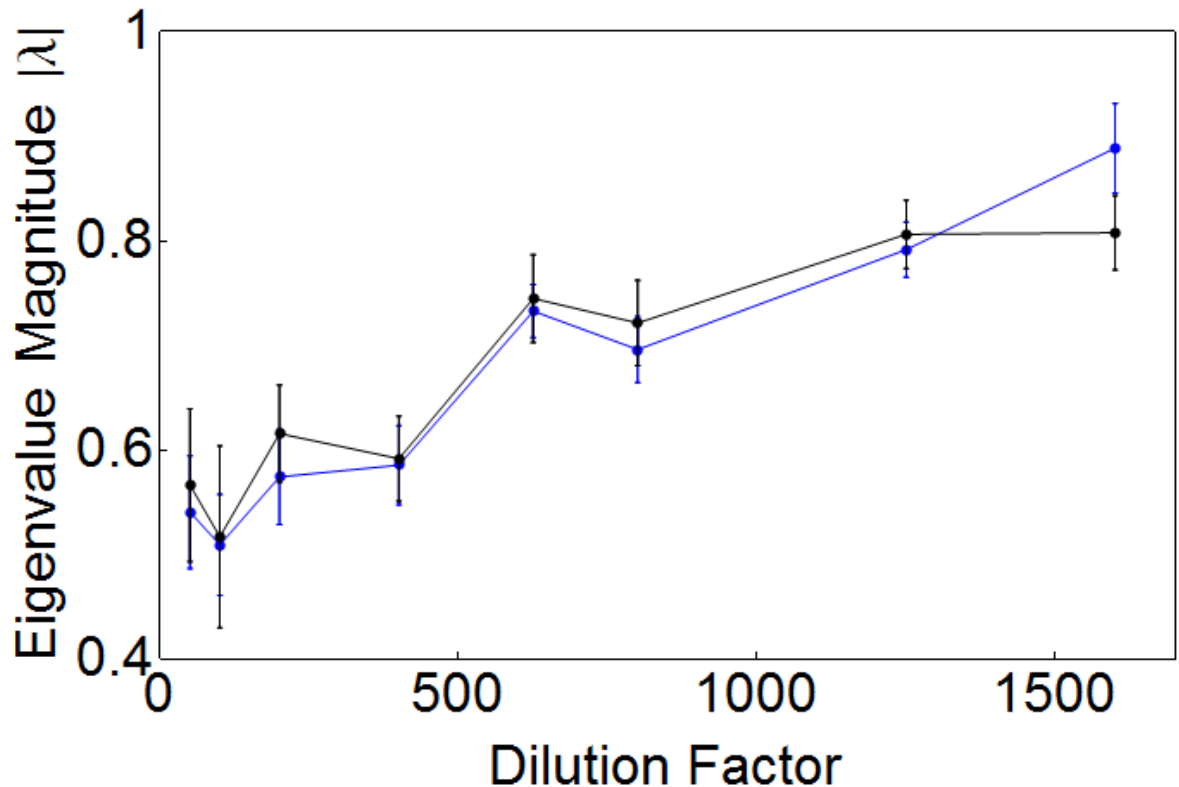

**Figure S3: Comparison of eigenvalue magnitude results from two different analysis methods.** The experiment described in Fig. 3 was analyzed in two different ways: a functional minimization of four parameters given a fixed point (Method 1, blue), and a parameter-fitting algorithm calculated over four parameters as well as the fixed point location (Method 2, black). The eigenvalue magnitudes calculated using these methods were plotted against the dilution factor. Error bars represent standard error, and were obtained through bootstrapping the trajectories included in the analysis, as well as the location of the fixed point for Method 1.

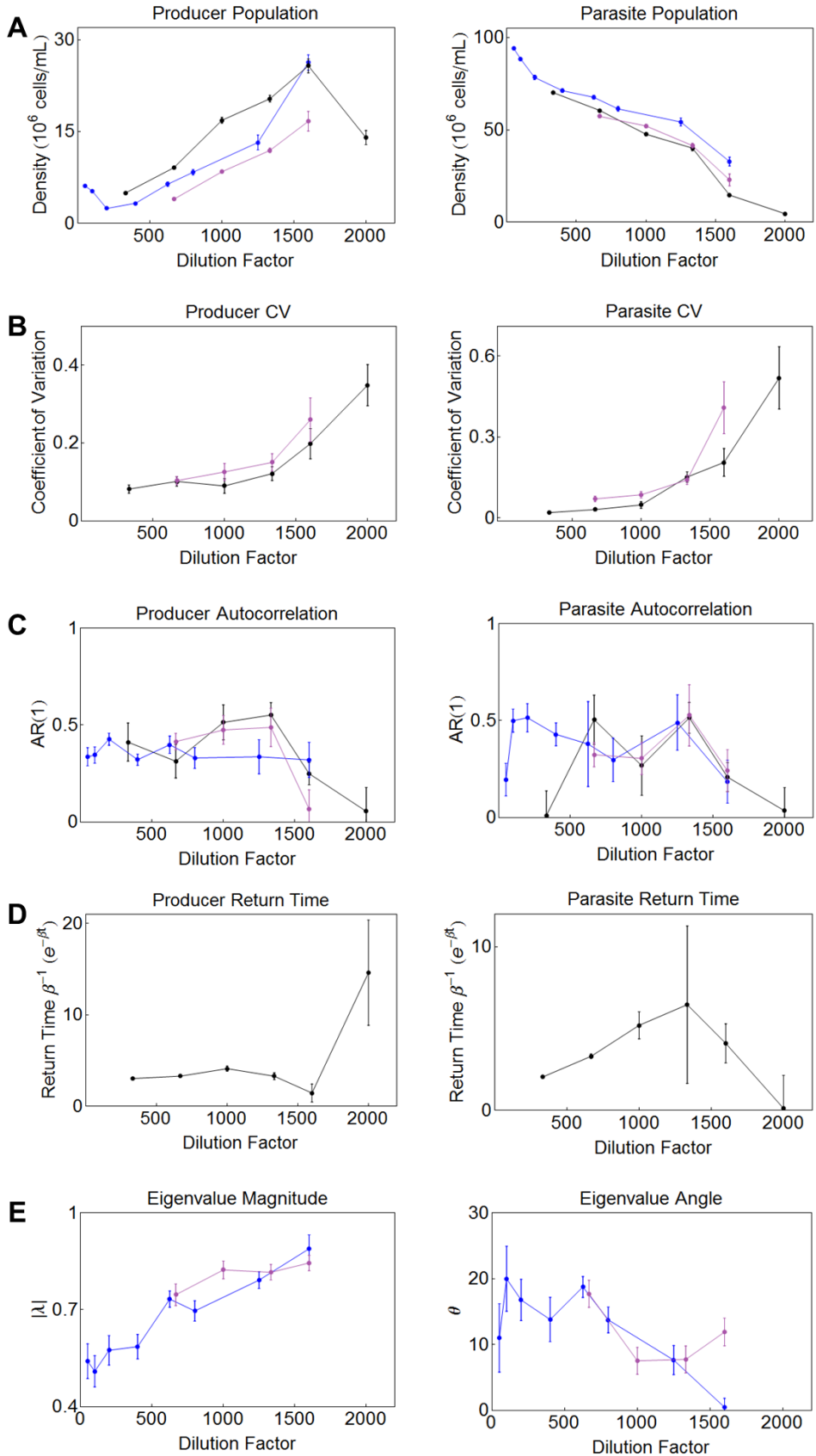

**Figure S4: Early warning indicators of ecosystem collapse.** A compilation of results from all of the three experiments for: **(A)** Mean Equilibrium Density for each population, **(B)** CV – Coefficient of Variation for each population, **(C)** *AR(1)* autocorrelation for each population, which was measured from Day 2 to the last day, **(D)** Return time (see Methods), and **(E)** $|\lambda|$ and $\theta$ - magnitude and argument of the eigenvalues. Data was filtered (see Methods) to exclude any obviously dying trajectories from the three experiments. Error bars represent standard errors. The Black data corresponds to experiment from Figs. 1 and 4, Violet to the one from Fig. 2, and Blue to the data from Fig. 3. Error bars represent standard errors and were attained through bootstrapping.

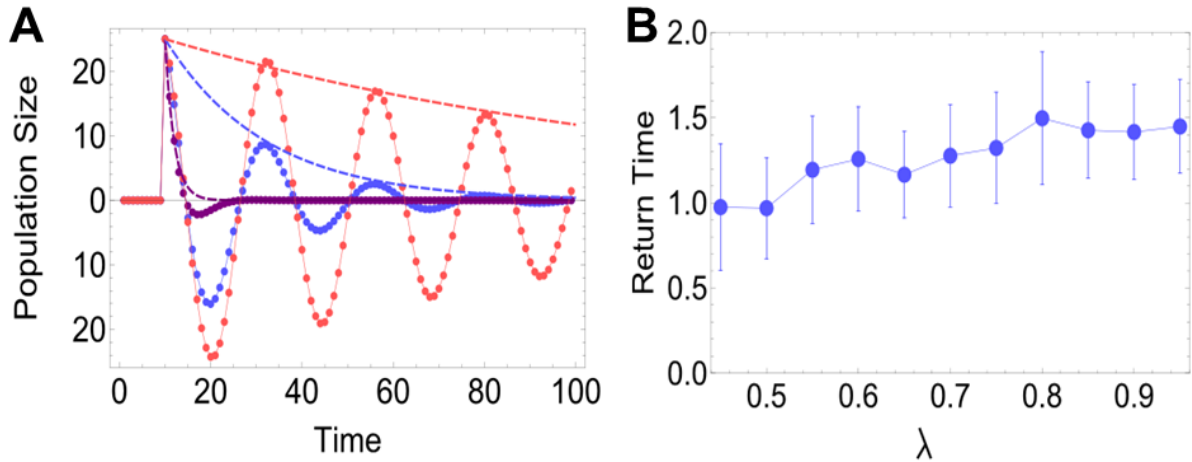

**Figure S5: Return to equilibrium after a perturbation as a function of the magnitude of the eigenvalue.** **(A)** We show the dynamical response of a population whose dynamics are governed by equation (7) (only one of the populations, *x*, is plotted), with a Jacobian matrix with imaginary eigenvalues (see equation (22)). The dynamics are characterized by damped oscillatory behavior, and thus they are described by two timescales; (i) the timescale of oscillations and (ii) the decay of the envelope of the oscillations. For simplicity we just plot the deterministic dynamics (noise strength is assumed to be 0). The orange dots correspond to $|\lambda|=0.99$, blue dots correspond to $|\lambda|=0.95$, and purple correspond to $|\lambda|=0.75$. All other parameters ($\theta=15^o, Q=45^o, r=1$) are kept constant. We find that while the envelope of the fluctuations (dashed lines) decays ever more slowly as $|\lambda|\to 1$, the short term decay is little sensitive to the value of $|\lambda|$. Both time and Return Time units are Days. **(B)** We repeated this simulation in the presence of noise ($\sigma=1$). The return time was naively estimated by fitting the short-term decay (over the first eight days) to an exponential decay function (mimicking what an unsuspecting researcher might do with experimental data collected in the field). As expected, we found that this "naïve estimate" of the decay time scales only weakly with $|\lambda|$, since it is the envelope what should decay more and more slowly as $|\lambda|$ increases, while the short term dynamics are mainly governed by the complex argument of the eigenvalue ($\theta$). Error bars represent standard deviation of the list of return times obtained in 50 simulations.

**Supplementary References**

1.  Scheffer M et al. (2009) Early-warning signals for critical transitions. *Nature* 461:53–59.

2.  Scheffer M et al. (2012) Anticipating critical transitions. *Science* 338:344–348.

3.  Scheffer M, Carpenter S, Foley JA, Folke C, Walker B (2001) Catastrophic shifts in ecosystems. *Nature* 413:591–596.

4.  Strogatz S (1994) *Nonlinear dynamics and chaos* (Westview Press).

5.  Van Nes EH, Scheffer M (2007) Slow recovery from perturbations as a generic indicator of a nearby catastrophic shift. *Am Nat* 169:738–747.

6.  Chisholm RA, Filotas E (2009) Critical slowing down as an indicator of transitions in two-species models. *J Theor Biol* 257:142–149.

7.  Dakos V, van Nes EH, D'Odorico P, Scheffer M (2012) Robustness of variance and autocorrelation as indicators of critical slowing down. *Ecology* 93:264–271.

8.  Dakos V et al. (2012) Methods for detecting early warnings of critical transitions in time series illustrated using simulated ecological data. *PLoS ONE* 7:e41010.

9.  Carpenter SR, Brock WA, Cole JJ, Kitchell JF, Pace ML (2008) Leading indicators of trophic cascades. *Ecology Letters* 11:128–138.

10. Drake JM, Griffen BD (2010) Early warning signals of extinction in deteriorating environments. *Nature* 467:456–459.

11. Dai L, Vorselen D, Korolev KS, Gore J (2012) Generic indicators for loss of resilience before a tipping point leading to population collapse. *Science* 336:1175–1177.

12. Veraart AJ et al. (2012) Recovery rates reflect distance to a tipping point in a living system. *Nature* 481:357–359.

13. Downing AS, van Nes EH, Mooij WM, Scheffer M (2012) The resilience and resistance of an ecosystem to a collapse of diversity. *PLoS ONE* 7:e46135.

14. Biggs R, Carpenter SR, Brock WA (2009) Turning back from the brink: detecting an impending regime shift in time to avert it. *Proc Natl Acad Sci USA* 106:826–831.

15. Thrush SF et al. (2009) Forecasting the limits of resilience: integrating empirical research with theory. *Proc Biol Sci* 276:3209–3217.

16. Lade SJ, Gross T (2012) Early warning signals for critical transitions: a generalized modeling approach. *PLoS Comput Biol* 8:e1002360.

17. May RM (2001) *Stability and complexity in model ecosystems* (Princeton University Press).

18. Ives AR, Dennis B, Cottingham KL, Carpenter SR (2003) Estimating community stability and ecological interactions from time-series data. *Ecological Monographs* 73:301–330.

19. Carpenter SR et al. (2011) Early warnings of regime shifts: a whole-ecosystem experiment. *Science* 332:1079–1082.

20. Gore J, Youk H, van Oudenaarden A (2009) Snowdrift game dynamics and facultative cheating in yeast. *Nature* 459:253–256.

21. Sanchez A, Gore J (2013) Feedback between population and evolutionary dynamics determines the fate of social microbial populations. *PLoS Biology* 11:e1001547.

22. Nowak MA (2006) *Evolutionary Dynamics* (Harvard University Press).

23. Kéfi S, Dakos V, Scheffer M, Van Nes EH, Rietkerk M (2012) Early warning signals also precede non-catastrophic transitions. *Oikos*:EV1–EV8.

24. Ives AR (1995) Measuring Resilience in Stochastic Systems. *Ecological Monographs* 65:217–233.

25. Kuznetsov Y (2004) *Elements of applied bifurcation theory* (Springer). 3rd Ed.

26. Boerlijst MC, Oudman T, de Roos AM (2013) Catastrophic Collapse Can Occur without Early Warning: Examples of Silent Catastrophes in Structured Ecological Models. *PLoS ONE* 8:e62033.

27. Hastings A, Wysham DB (2010) Regime shifts in ecological systems can occur with no warning. *Ecol Lett* 13:464–472.

28. Boettiger C, Hastings A (2012) Early warning signals and the prosecutor's fallacy. *Proc Biol Sci* 279:4734–4739.

29. Boettiger C, Hastings A (2013) Tipping points: From patterns to predictions. *Nature* 493:157–158.

30. Neubert MG, Caswell H (1997) Alternatives to resilience for measuring the responses of ecological systems to perturbations. *Ecology* 78:653–665.

31. May RM (1977) Thresholds and breakpoints in ecosystems with a multiplicity of stable states. *Nature* 269:471–477.

32. Dai L, Korolev KS, Gore J (2013) Slower recovery in space before collapse of connected populations. *Nature* 496:355–358.

33. Celiker H, Gore J Competition between species can stabilize public-goods cooperation within a species. *Molecular Systems Biology* In press.